\documentclass[manuscript]{aastex}

\slugcomment{To be submitted to the Astronomical Journal}

\shorttitle{Spatial density distribution of stars around four globular clusters in the Galactic bulge region}
\shortauthors{Chun et al.}

\begin{document}

\title{Tidal stripping stellar substructures around four metal-poor globular clusters in the Galactic bulge}

\author{Sang-Hyun Chun\altaffilmark{1}, Minhee Kang\altaffilmark{2}, DooSeok Jung\altaffilmark{2}, Young-Jong Sohn\altaffilmark{2}}

\altaffiltext{1}{Yonsei University Observatory, Seoul, Korea; shchun@galaxy.yonei.ac.kr}
\altaffiltext{2}{Department of Astronomy, Yonsei University, Seoul 120-749, Korea}

\begin{abstract}
We investigate the spatial density configuration of stars around four metal-poor globular clusters
(NGC 6266, NGC 6626, NGC 6642 and NGC 6723) in the Galactic bulge region using wide-field 
deep $J$, $H$, and $K$ imaging data obtained with the WFCAM near-infrared array on United Kingdom 
Infrared Telescope. 
Statistical weighted filtering algorithm for the stars on the color-magnitude diagram is applied in order to sort 
cluster member candidates from the field star contamination. 
In two-dimensional isodensity contour maps of the clusters, we find that
all of the four globular clusters exhibit strong evidence of tidally stripping stellar features beyond tidal radius, in the form of
tidal tail or small density lobes/chunk. The orientations of the extended stellar substructures are likely to be associated with 
the effect of the dynamic interaction with the Galaxy and the cluster's space motion. 
The observed radial density profiles of the four globular clusters also describe the extended substructures;
they depart from theoretical King and Wilson models and have an overdensity feature with a break in a slope of profile at the 
outer region of clusters.
The observed results could imply that four globular clusters in the Galactic bulge region have experienced 
strong environmental effect such as tidal force or bulge/disk shock of the Galaxy in the dynamical evolution of the globular clusters. 
These observational results provide us further constraints to understand the evolution of clusters in the Galactic bulge region as well as the formation of the Galaxy. 
\end{abstract}


\section{INTRODUCTION}
According to modern cold dark matter cosmology, galaxies are hierarchically assembled by the merging
or accretion of small fragments~\citep{Bau96,Kly99,Moo99,Diemand2007}. In this theory, the stellar
halo of galaxies, such as the Milky Way, are mostly built up from small substructures such as satellite 
galaxies~\citep{Searle1978,Joh98,Bul01,Aba06,Fon06,Moo06}. These satellite systems suffer significant 
tidal disruption and mass loss by the tidal force and shock of host galaxies during the process of accretion, 
thereby producing a number of stellar substructures, such as tidal tails or streams in the galactic halo~\citep{Bul05}.
Thus, the study of stellar streams in the Milky Way is a valuable for reconstructing the accretion history of the
Galaxy~\citep{Koposov2010,Law2009} and for understanding the potential of the Galaxy~\citep{Ode09} 

In the last two decades, numerous stellar streams and tidal tails have been discovered in the Galactic halo.
The Sagittarius dwarf galaxy and its stellar streams~\citep{Iba94,Iba95,Iba97,Iba01,Viv01,Maj03,New03,Mar04,Bel06a}
are the most well studied out of the many other recently discovered stellar streams
~\citep{Hel99,Ive00,Yan00,Yan03,New02,Mart04,Roc04,Mar05,Duf06,Gri06,Jur08}.
Sky survey projects, such as the Sloan Digital Sky Survey (SDSS) and the Two Micron All Sky Survey (2MASS),
are discovering more stellar substructures in the Galactic halo; newly-discovered stellar streams include
the Virgo stellar stream~\citep{Vivas2006, Vivas2008}, the Orphan Stream~\citep{Gri06,Zuc06,Bel07}, and 
the Cetus stream~\citep{New09, Koposov2012}. More recent works have reported that some of these 
streams  are associated with globular clusters~\citep{Drake2013,Grillmair2013}.

Globular clusters have been one of the most investigated stellar systems. They have provided crucial information 
about the formation and evolutionary mechanisms of the Galaxy. However, recent photometric and spectroscopic studies are still
amending the accepted the view of  how the globular clusters formed and their contribution to the formation of the Milky Way.
It appears that they are not just simple stellar populations, as was previously thought~\citep{Gratton2004,Carretta2009}, 
and some of them, such as $\omega$ Centauri~\citep{Lee99} and NGC 6656~\citep{Lee09},
are even considered to be surviving remnants of the first building block that merged into the Milky Way.
Several globular clusters in the Milky Way~\citep[about 27\% of the Milky Way's globular clusters;][]{Mac04} could have
formed via the accretion or merging of more complex systems. In addition, recent work suggests that globular clusters 
were $8-25$ times more massive than they are present when they first formed~\citep{Conroy2011,Schaerer2011}.
Thus, the stellar streams around globular clusters are important objects to study for understanding the merging or accretion 
history of the Milky Way and to gather information regarding the dynamical evolution of globular clusters.
Indeed, the remarkable long tidal tail of  Palomar 5~\citep{Ode01,Gril06a} and NGC 5466~\citep{Bel06b,Grill06}, as well as
the presumed globular cluster stream GD-1 ~\citep{Gril06b}, are
spectacular examples of globular cluster streams. 
The tidal bridge-like features and common envelope structures around M53 and NGC 5053~\citep{Chu10} are also particularly
interesting, as they are the evidence of an accretion event of dwarf galaxies into the Milky Way.
Slightly extended tidal substructures also appear in the vicinity of several globular clusters~\citep{Gri95,Leo00,Soh03}.

Despite numerous discoveries of globular cluster streams, most of the globular cluster streams found to date have been
in the Galactic outer halo. However, there are more than 40 globular clusters in the Galactic bulge region,
and the origin of metal-poor globular clusters in the bulge region is still unclear. 
There have been a few studies of the stellar streams of the globular clusters in the bulge region.
The stellar substructure around globular cluster NGC 6626 was the first discovery in the bulge region~\citep{Chun2012}.
In a hierarchical model, it can be seen that vigorous merging events of subclumps 
exist in the bulge region of galaxies like the Milky Way~\citep{Kat92,Bau96,Zoc06}. These merging events then result in a wide
metallicity distribution~\citep[$-1.5\leq\lbrack Fe/H\rbrack<0.5$;][]{McW94,Zoc03} of stars in the bulge region~\citep{Nak03}.
Terzan 5 is an example of a merging event in the bulge region in the past~\citep{Fer09}.
Therefore, we can expect to discover extratidal substructures around some of the globular clusters in the
bulge region.

In this study, we investigated the spatial density distribution of stars around four metal-poor globular clusters
in the Galactic bulge region - NGC 6266, NGC 6626, NGC 6642, and NGC 6723.
We assigned the bulge region as the area within $3$ kpc from the Galactic center.
Table ~\ref{para} shows the basic parameters of the four globular clusters.
In order to reduce the effect of high extinction toward the bulge,
we used wide-field ($45'\times45'$) near-infrared $JHK$ photometric data obtained 
from the observation of the Wide Field Camera (WFCAM) array attached to the United Kingdom Infrared Telescope (UKIRT).
Section 2 presents our observations, data reduction process, and photometric measurements.
The statistical analysis and filtering technique used for member star selection are described in Section 3. 
In Section 4, we investigate two-dimensional stellar density maps and the radial profile of the clusters 
to trace the stellar density features. 
The discussion of our investigation is presented in Section 5.
Lastly, we summarize the results and discussion in Section 6.

\section{OBSERVATION, DATA REDUCTION AND PHOTOMETRY}
Photometric imaging data for four globular clusters were observed using the WFCAM on the 3.8 m 
UKIRT in Hawaii in April and July of 2010. The WFCAM is an infrared mosaic detector of four Rockwell 
Hawaii-II (HgCdTe 2048$\times$2048) arrays with $12^{'}.83$ gap between the arrays.
Four separately pointed observations (four tiles) result in a filled-in sky area of 0.75 square degrees with a pixel scale of $0'.4$. 
Our target clusters were observed using the four-tile observations in three band filters ($J, H$, and $K$) to get continuous sky images covering
a total field-of-view of 0.75 square degrees.
The individual image of each cluster for one tile was recorded in short ($1$ sec for $JHK$) and long ($5$ sec for $JH$, and $10$ sec for $K$) exposures to optimize the photometry of bright and faint stars. A five-point dithering pattern was applied to reject bad pixels and cosmic rays.
At each dithered position, $2\times2$ micro-stepping observation was also carried out to get well-sampled stars.
A separate sky observation was obtained for removing thermal background emission after observing the target images.
We also observed several comparison fields in the bulge region using the same observation strategy for observing the clusters.
The stars in the comparison field area were used in the processes of color-magnitude (C-M) mask filtering technique and the optimal contrast filtering technique in the following section in order to estimate the field star contamination around the globular clusters on the
color-magnitude diagram (see Section 3).
Three comparison fields were finally selected using the following condition: the comparison field was not very distant from the clusters on the sky, and the morphology
of the color-magnitude diagram for the field stars was similar to that of the globular clusters.
The coordinates of the selected comparison fields  were indicated in Table ~\ref{para}.
Table ~\ref{log} provides the exposure time of each filter for the four 
globular clusters.

Standard data reduction for near-infrared imaging, which includes dark subtraction, flat fielding, and the removal of
crosstalk, was completed by the pipeline of the Cambridge Astronomy Survey Unit (CASU). 
Then thermal emission backgrounds were made by median-combining the CASU-processed images of the separate sky observations.
The resulting blank sky images were subtracted from the all target images. The residual sky background level of each target
image was also removed in each target image. All sky-subtracted images were interleaved into a single image for photometric
analysis using Swarp~\citep{Bertin2002}. 
The final resampled images of the four globular clusters have a wide-field area of about $45'\times45'$, which is sufficiently 
large area to cover from the center of each target cluster to two times its tidal radius.
The average seeing condition of stars in the resampled images is between $0.75\sim1.05$ arcsec. Table~\ref{log}
summarizes the average FWHM values for each filter.

Stellar photometry on each detector was performed using the point-spread function (PSF) fitting routine ALLSTAR ~\citep{Stetson1988}.
The PSF varying quadratically through position was first constructed with
$100-150$ bright and isolated stars using the DAOPHOT II program~\citep{Stetson1987}.
The quality of the PSF was improved by removing the neighboring faint stars around the PSF stars and iteratively reconstructing the PSF.
Then, the instrumental magnitude of the individual stars on each array was estimated by the ALLSTAR process using improved PSF. 
The raw positions of the stars on the detector were transformed into an equatorial coordinated system using the Two Micron 
All Sky Survey (2MASS) point-source catalog. The instrumental magnitudes of the stars were transformed on to the 2MASS filter system
using the color term between WFCAM and the 2MASS system~\citep{Dye2006}. Then the photometric zero-points were finally
computed and calibrated by comparing the magnitudes of common stars in our photometric catalog and 2MASS catalog.
The astrometric and photometric data of each chip on a mosaic were finally combined into 
a whole set of data for the target cluster. Stellar objects with photometric measurement error larger than $0.1$ mag were removed
in order to reduce the spurious detection. 
We also measured the individual extinction value of each star according the position of the 
sky by using the map of ~\citet{Sch98}; the mean E(J-K) and extinction values in $K$ are  
$E(J-K)=0.19$ and $A_K=0.14$ for NGC 6266,
$E(J-K)=0.24$ and $A_K=0.18$ for NGC 6626,
$E(J-K)=0.20$ and $A_K=0.15$ for NGC 6642, and
$E(J-K)=0.12$ and $A_K=0.08$ for NGC 6723.
We subtracted derived extinction values from the observed magnitude.

\section{PHOTOMETRIC FILTERING FOR MEMBER STAR SELECTION}
In order to accurately trace the stellar distribution around the globular clusters, 
it is important to reduce the contamination of the field stars and to enhance
the density contrast between the cluster candidate stars and the field stars.
Although many statistical methods for filtering field stars have been introduced in the past few decades, 
the color-magnitude (C-M) mask filtering technique ~\citep{Gri95} and the optimal contrast filtering technique ~\citep{Ode03}
were frequently used. We also basically followed these two methods ~\citep[for a detailed description, see][]{Gri95,Ode03,Chun2012}.

We first define new orthogonal color indices $c_1$ and $c_2$ from the one-dimensional distribution of stars in a $(J-K)$ versus $(J-H)$
color-color diagram~\citep[see Figure 2 of][]{Chun2012}. The color indices were chosen in such a way that the $c_1$ axis was placed along the main distribution of the stars, while the $c_2$ axis was perpendicular to the $c_1$ axis. Equation~\ref{eq:rela1} shows the general forms of two orthogonal color indices, and Table~\ref{coefficients} indicates the coefficients $a$ and $b$ of the new color indices for each cluster. 

\begin{eqnarray}
\label{eq:rela1}
c_1=a(J-K)+b(J-H), \\
c_2=-b(J-K)+a(J-H) \nonumber
\end{eqnarray}

\begin{figure*}
\centering
\includegraphics[width=0.6\textwidth]{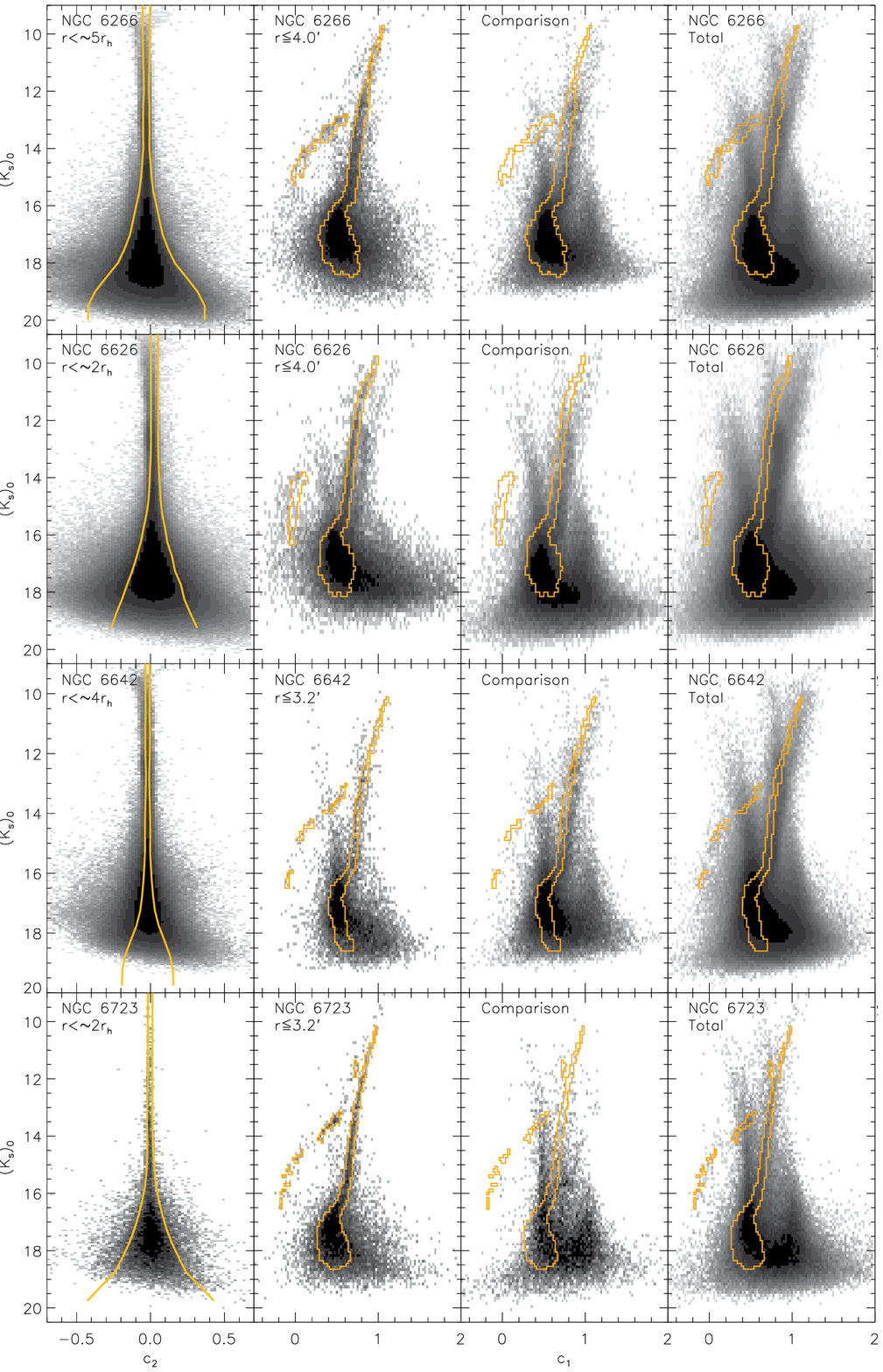}
\caption{$(c_2,K)$ and $(c_1,K)$ color magnitude diagrams of stars for four globular clusters. 
The left panel shows the $(c_2,K)$ CMD of stars in central region of clusters.
The lines in the $(c_2,K)$ plane are $2\sigma_{c_2}(K)$ rejecting limit 
at $K$ magnitudes.
The second, third and fourth panels are the $(c_1, K)$ CMDs for the stars
in the cluster central region, in the assigned comparison region, and in the total field of four clusters.
The grid lines in $(c_1,K)$ CMDs indicate the filtering mask envelope where the signal-to-noise ratios of cluster
star counts are maximized through a C-M mask filtering technique. The stars outside the grid lines are highly
unlikely to be cluster members. The optimal contrast filtering technique was also applied to this envelope.}
\label{c1c2cmd} 
\end{figure*}

In the $(c_2, K)$ color-magnitude diagram (CMD), we rejected all stars with $|c_2|>2\sigma_{c_2}(K)$, where $\sigma_{c_2}(K)$ is the dispersion
in $c_2$ for stars with magnitude $K$.
The stars within $2\sim5r_h$ for each cluster were used when we defined the rejecting limit.
The left panel of Figure~\ref{c1c2cmd} shows a $(c_2, K)$ CMD for stars in $2\sim5r_h$ from the cluster center. 
The lines indicate our rejecting limit; we considered that stars outside this boundary were 
unlikely to be cluster member stars.

After this preselection in the $(c_2, K)$ plane, 
we defined the locus of cluster on the CMD where the signal-to-noise ratio (S/N) of the cluster star count was maximized in contrast to the comparison field stars using the C-M mask filtering technique in the $(c_1, K)$ plane.
First, we made a representative sample of CMDs for the cluster and the fields using the stars in the central region of the cluster and the observed comparison fields.
The second and third panels of Figure~\ref{c1c2cmd} 
show the $(c_1, K)$ CMDs of stars within $3'.0\sim4'.0$ from each cluster center and
the selected comparison region, respectively.  The right panel shows the $(c_1, K)$ CMD of the stars in the total
survey region for the cluster. Then, the CMDs of the cluster and comparison were subdivided into small
subgrid elements, and the signal-to-noise ratio in each subgrid element was calculated using Equation~\ref{eq:sn1}:
\begin{eqnarray}
\label{eq:sn1}
s(c_1,K)=\frac{n_{cl}(c_1,K)-gn_f(c_1,K)}{\sqrt{n_{cl}(c_1,K)+g^2n_f(c_1,K)}},
\end{eqnarray}
where $n_{cl}(c_1,K)$ and $n_{f}(c_1,K)$ are the number of stars in the subgrid elements for the cluster and comparison region, respectively;
$g$ is the area ratio of the cluster region to comparison region.
From array $s$, we computed the cumulative number of stars for the cluster $N_{cl}(k)$ and comparison $N_f(k)$, respectively, 
by sorting the elements of $s(c_1,K_s)$  into a series of descending order with a one-dimensional index of $k$.
Then, a cumulative signal-to-noise ratio $S(k)$ was calculated by Equation~(\ref{eq:csn1}):
\begin{eqnarray}
\label{eq:csn1}
S(k)=\frac{N_{cl}(k)-gN_f(k)}{\sqrt{N_{cl}(k)+g^2N_f(k)}}
\end{eqnarray}
 $S(k)$ become a maximum value for a specific subarea of the C-M plane, and the $s(c_1, K)$ corresponding to the maximum
 value of $S(k)$ was chosen as an optimal threshold, $s_{lim}$. 
The filtering mask area in the $(c_1,K)$ plane was then determined by selecting subgrid elements
with larger $s(c_1,K)$ values than the determined $s_{lim}$.
The solid lines in the second, third, and fourth panels in Figure~\ref{c1c2cmd}
represent the selected filtering mask envelope.
The entire sample of stars in the determined filtering mask
area was considered in the following filtering analysis.

Finally, we applied the optimal contrast filtering technique to the stars in the determined filtering mask envelope obtained from the C-M mask
filtering technique. We calculated the number density distribution of star in $(c_1, K)$ C-M plane (Hess diagram)
for cluster and comparison field. The bin size of the Hess diagram is the same as that of the C-M mask filtering technique. 
Then, the density of cluster stars $n_c(k)$ at a given position $k$ on the sky was derived by Equation~\ref{eq:optimal}:
\begin{eqnarray}
\label{eq:optimal}
n_{c}(k)=\frac{\sum_j[n(k, j)f_c(j)/f_F(j)-n_F(k, j)f_c(j)/f_F(j)]}{\sum_jf^2_c(j)/f_F(j)},
\end{eqnarray}
where $n_c(k, j)$ and $n_F(k, j)$ are cluster star density and field star density in the $j$th subgrid in the 
optimal mask envelope of the C-M plane and in the $k$th bin in position on the sky; $f_c$ and $f_F$ are the 
normalized density distribution of the cluster and comparison field in the optimal mask envelope in the Hess diagram of $(c_1, K)$.
In the optimal contrast filtering technique, the ratio $ f_c(j)/f_F(j)$ of the number density distribution of the cluster stars to the comparison field 
stars in optimal mask envelope of $(c_1, K)$ CMD was used as conditional weight to determine cluster membership.
The number density of stars in the sky was calculated by summing up the conditional weights of all stars and dividing this
sum by the factor $a=\sum_jf^2_c(j)/f_F(j)$. Thus, this returns the estimated number of cluster stars $n_c$ plus a term of $n_F/a$,
the number of contaminating field stars attenuated by $a$.

\section{Spatial density features of stars in the vicinity of the four globular clusters}
In this section, we present the spatial density features of the stars in the vicinity of the four globular clusters.
The large area of the WFCAM data ($45'\times45'$ of the sky) enable us to examine the features of the stellar 
density distribution from the cluster center to a distance of at least two times the tidal radius.
The two-dimensional density distribution and the radial density profile for each cluster were investigated using the 
selected stars by C-M mask filtering technique and the weighted number obtained from the optimal contrast filtering technique.

\begin{figure}
\centering
\includegraphics[width=1.0\textwidth]{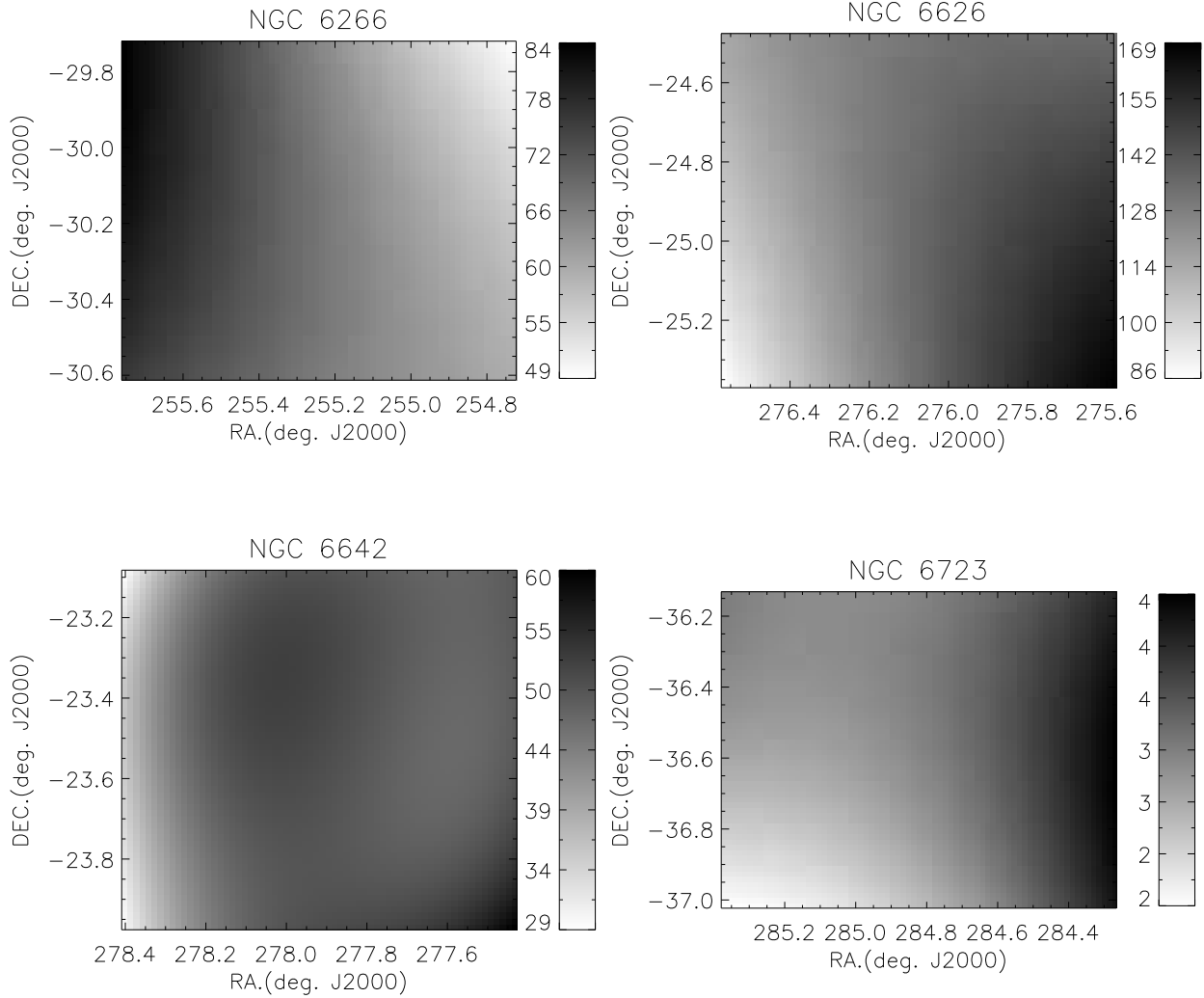}
\caption{The field star contribution maps for four globular clusters. The density gradients and variation across
the globular clusters were described with gray scale. The sidebar indicates the number of stars per square pixels.
}  
\label{background} 
\end{figure}

The two-dimensional stellar surface density maps of the clusters were constructed using Equation~\ref{eq:optimal}.
The sky plane of cluster was divided into small grids with pixel width of ~$0'.9\times0'.9$, and the weighted 
counts of stars were calculated in those pixels. 
The field stars contamination was then constructed by masking the central region within $1\sim1.5r_t$ and fitting
a low-order bivariate polynomial model. Figure~\ref{background} shows the constructed field stars contamination for each
cluster, and the density gradients or variations of field stars across the globular cluster were represented with gray scale.
We subtracted these field stars contamination maps and made the field across the
globular cluster essentially flat. The residual background density map was also made using the same method with field
star contamination maps, but in this case, we just
subtracted the mean density level of residual background density map.
The star number density map of each cluster was then smoothed with a Gaussian smoothing algorithm to
increase signal-to-noise ratio and enhance the spatial frequencies of interest.
The isodensity level was described by contour with a standard deviation unit $(\sigma)$ of the 
background level on the smoothed map with the various kernel values. 
The distribution map of the E(B-V) value for the observed region was also derived from the 
map of ~\citet{Sch98} to examine possible extinction effects.

The radial number density profiles of the globular clusters are useful for understanding the internal and outer structure of the globular clusters. 
This overall structure has been described for a long time by~\citet{King66} model, which is characterized by a truncated density
profile at the outer edge. However, according to the results of a recent wide-field observation, 
the radial number density profiles of several globular clusters are not truncated at their outer edges; instead, they have an extended 
overdensity feature that departs from the behavior predicted by~\citet{King66} model and smoothly drops toward
the background level~\citep{Gri95,Leo00,Tes00,Roc02,Lee03,Ols09, Carballo2012}.
Numerical simulations also reproduced and characterized this overdensity feature as a break in the slope of 
radial profile due to the extratidal stars around globular clusters~\citep{Com99,Joh99,Joh02}.
In these models, the break in the slope of the radial profile was described by the power law $r^{-\gamma}$.
~\citet{Wilson1975} models has also been used to describe the structure of globular clusters.
Indeed,~\citet{McLau05} fitted Wilson model to the radial density structure of globular clusters in the Galaxy and the 
Magellanic Clouds. We note that Wilson model is spatially more extended than King model~\citep[see][]{McLau05}.

\begin{figure}
\centering
\includegraphics[width=0.4\textwidth]{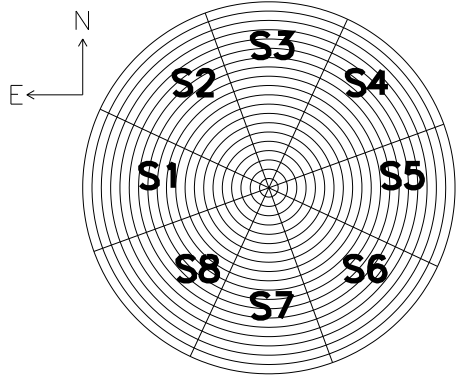}
\caption{Reseau plot used for radial density profile. The radial surface densities were measured in
concentric annuli. We divided each annulus into eight angular sections (S1-S8) in order to derive the surface 
density profiles in a different direction.}
\label{annulus}
\end{figure}

We also derived the radial surface density profile of each cluster and tried to find
the evidence of an extratidal extension at the outer edges of the clusters.
In order to construct the radial density profiles, we used concentric annuli with a width of $0^{'}.45$ ranging
from the cluster center out to a radius of $20^{'}.0$, and then counted the weighted number of stars in each annulus.
The number density of stars was then calculated by dividing the sum by the area of the annulus.
The field stars contribution on the radial profile was estimated from the field star contamination maps in Figure~\ref{background}, and
subtracted from the radial profile. Then the residual background density level which was measured on the residual background density map was also removed from the counts.
The error on the number density was estimated by the error propagated from a Poisson statistic for star counts.
We examined the radial completeness for measuring the crowding effect of the inner regions of the clusters by applying the artificial star test 
and then recovered the crowding effect using the radial complete ratio. 
However, the most central regions of the clusters were not resolved enough to derive the 
number density profile because of the crowding effect even though we compensated for the number density. Therefore, for the central regions, 
we combined our number density profiles with the previously published surface brightness profiles of ~\citet{Tra95}.
The surface brightness profile of ~\citet{Tra95} was converted into a number density scale by the equation, $log N(r)=-\mu(r)/2.5+C$, where
$C$ is an arbitrary constant to match the number density profile and surface brightness profile.
Then, the final number density profiles were empirically fitted by King model and Wilson model.
We also derived radial surface density profiles for a different direction,
for which we divided an annulus into eight sections (S1-S8)
with an angle of $45^{\circ}$, as shown in Figure~\ref{annulus}.
The annulus widths were assigned to be  $\sim0'.5$ at innermost region with $r<5'$, $\sim1'.0$ at the middle region
with $5'<r<10'$, and $2'.0$ at outer region with $10' < r < 20'$. 
We note that some radial density points with small number statistics could not be plotted 
because the number densities in these regions were lower than background density level.

\subsection{NGC 6266 (M62)}
\begin{figure}
\centering
\includegraphics[width=0.5\textwidth]{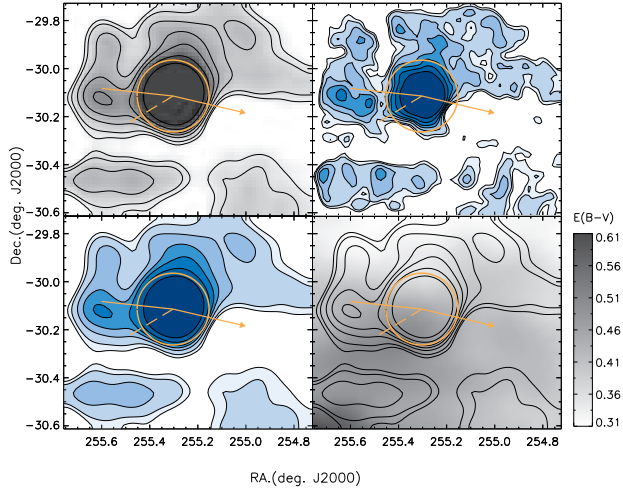}
\caption{From top-left to bottom-right, the star count map around NGC 6266, the surface
density maps smoothed by Gaussian kernel values of $0^{\circ}.045$ and $0^{\circ}.12$, overlaid with 
isodensity contour levels $0.5\sigma, 1.0\sigma, 2.0\sigma, 3.0\sigma, 4.0\sigma, 6.0\sigma$ and $10.0\sigma$, 
and the distribution map of E(B-V).
The circle indicates a tidal radius of NGC 6266. The arrow represent the proper motion of cluster.
Solid line indicates the direction of the Galactic center, and dashed line shows the perpendicular direction of
the Galactic plane.}
\label{N6266contour} 
\end{figure}

The star count map around NGC 6266 and the surface density maps smoothed by a Gaussian kernel value of
$0^{\circ}.045$ and $0^{\circ}.12$ are shown in Figure~\ref{N6266contour}, from the top-left panel to
the bottom-left panel. The grey density map in the bottom-right panel of Figure~\ref{N6266contour}
is the distribution map of E(B-V) of ~\citet{Sch98}. The contour lines indicate $0.5\sigma, 1.0\sigma, 2.0\sigma, 3.0\sigma, 4.0\sigma, 
6.0\sigma$ and $10.0\sigma$. The contour lines with a Gaussian kernel value of
$0^{\circ}.12$ are overlaid in a star count map and the E(B-V) map. The direction toward the Galactic center and the perpendicular 
direction to the Galactic plane are indicated as a solid line and dashed line, respectively. The proper motion of NGC 6266, i.e.
$\mu_{\alpha}\cos\delta=-3.50\pm0.37$ mas yr$^{-1}$ and $\mu_{\delta}=-0.82\pm0.37$mas
yr$^{-1}$ ~\citep{Din2003}, is indicated by a long arrow.
The circle in each panel is the tidal radius of $r_t=8'.97$ ~\citep{Har96}.
 
Figure~\ref{N6266contour} clearly shows overdensity substructures around NGC 6266 that extend toward east and north-west
directions to $\sim1.5r_t$ at levels larger than $0.5\sigma$. 
The stellar substructure in the east direction lies along the direction of the Galactic center and the opposite direction of the proper motion.
In addition, the extended structure to the north-west is likely aligned with the opposite perpendicular direction to the Galactic
plane, and its marginal extension seems to bend to the direction of proper motion.
We note that the density feature at the southern region of the cluster is likely to be affected by the dust extinction as shown in bottom-right panel
of Figure~\ref{N6266contour}. 

\begin{figure}
\centering
\includegraphics[width=0.47\textwidth]{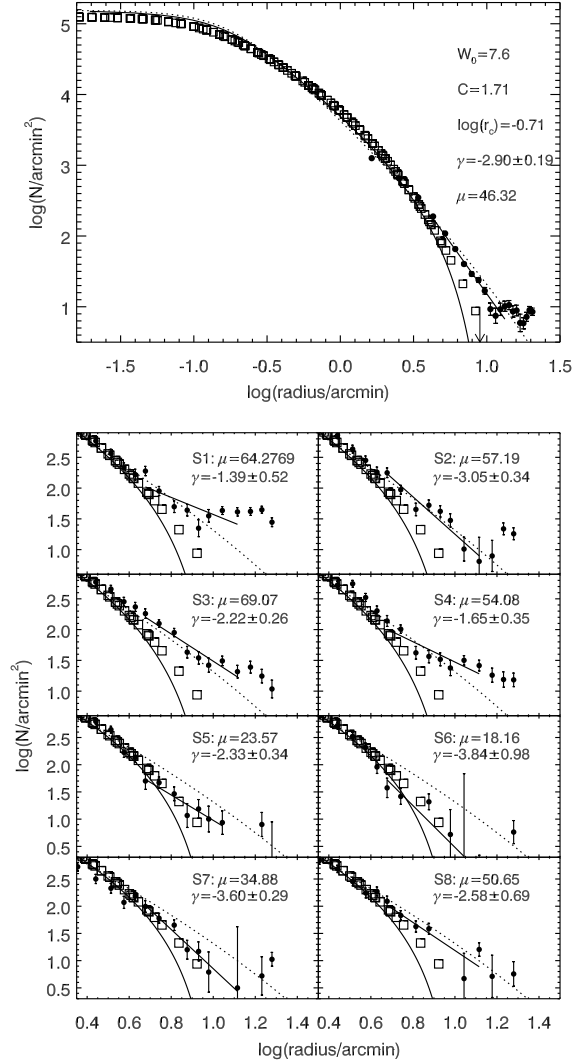}
\caption{{\it Upper} : Radial surface density profile of NGC 6266 with a theoretical King
model (solid curve) and Wilson model (dotted curve).
In central region, we plot surface brightness profile~\citep{Tra95} as open squares.
The arrow indicates the tidal radius ($r_t=8'.97$) of NGC 6266.
The overdensity feature at the region of $\sim 0.5 r_t\leq r\leq \sim 1.5 r_t$ was described
by a power-law, i.e., a straight line in logarithmic scale
with a slope of $\gamma=-2.90\pm0.19$.
The mean number density of stars in the overdensity region is estimated
as $\mu=46.32$ per arcminute square.
{\it Lower} : Radial surface density profiles of eight different angular
sections (S1$\sim$S8). The other notations are the same as those of the
radial surface density profile in the upper panel.}  
\label{N6266radial} 
\end{figure}

In the upper panel of Figure~\ref{N6266radial}, the radial surface density profile of NGC 6266 is plotted,
along with King model and Wilson model, which are arbitrarily normalized to our measurements.
In the central region of the cluster, we replaced the number density profile with the surface brightness of~\citet{Tra95}.
The profile of ~\citet{Tra95} connects smoothly with our number density profile at a radius of $log(r')\sim0.4$.
However, the number density profile shows an overdensity feature which departs from the King model and profile of ~\citet{Tra95},
with a break in the slope at the radius of log($r'$)$\sim0.65$ ($\sim0.5r_t$).
Here, we note that the profile of~\citet{Tra95} in the outer region might suffer from the background contamination 
and biases of bright stars~\citep[see][]{Chun2012,Noy06}, while our number density profile in this study has no bias due to bright stars.
The overdensity feature seems to extend to the radius of  log($r'$)$\sim1.15$ ($\sim1.5r_t$).
Wilson model shows a more extended profile to the outer region and seems to fit better with our measurements than the King model.
The excess density at this radial distance resembles a radial power law with a slope of $\gamma =-2.90\pm0.19$, which is
steeper than the case of $\gamma = -1$, predicted for a constant mass-loss rate over a long time ~\citep{Joh99}.
Thus, the overdensity feature at the outer region of NGC 6266 is indeed evidence of the extended substructures shown in Figure~\ref{N6266contour}.

The lower panel of Figure~\ref{N6266radial} shows the radial surface number density profiles for eight angular sections 
with a different direction as shown in Figure ~\ref{annulus}. 
We note that the some radial density points in section 5,6,7, and 8 were
not presented because the number densities in these
region were lower than subtracted residual background level.
The radial profiles in sections 1,2,3, and 4 show the overdensity features at the radius of $0.5r_t\lesssim r \lesssim 1.5r_t$.
The overdensity features in sections 1,3 and 4 seem to extend to more distant radius than $1.5r_t$.
The mean surface densities ($\mu$) in these sections are particularly 
higher than the total average density and those in the other sections.
In addition, the slope of profile in the sections 1 and 4 
are shallower than the mean slope of profile and those of other angular sections.
These mean density levels and shallow slopes in these sections are in
good agreement with the extended stellar substructures in the direction of the Galactic center, in the opposite direction 
of the proper motion, and in the opposite perpendicular direction to the Galactic plane.
On the other hand, the overdensity features do not appear in section 5,6, and 7 where prominent stellar substructures
were not shown in two-dimensional contour map.
The mean densities without overdensity features were also somewhat lower than total average density.


\subsection{NGC 6626 (M28)}
\begin{figure}
\centering
\includegraphics[width=0.5\textwidth]{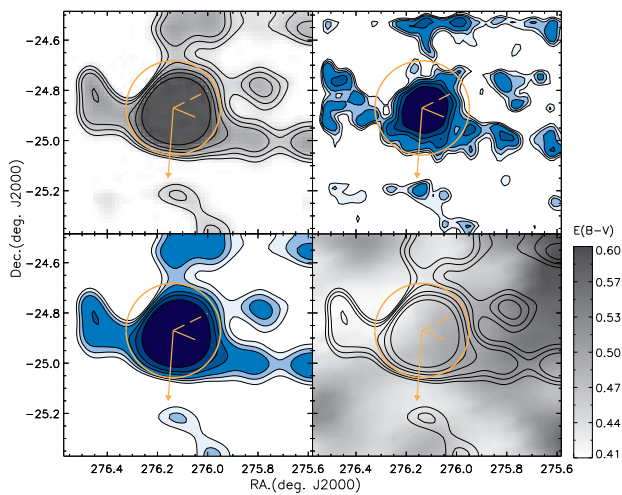}
\caption{The star count map around NGC 6626, surface density contour maps 
smoothed by Gaussian kernel values of $0^{\circ}.045$ and $0^{\circ}.12$, and distribution map of E(B-V) of~\citet{Sch98}
were plotted from top-left to bottom-right panel.
The contour levels indicate $1.0\sigma, 1.5\sigma, 2.0\sigma, 3.0\sigma, 4.0\sigma$, and $6.0\sigma$.
The long arrow indicated the proper motion of cluster.
The different lines indicate the direction of the Galactic center(solid line) and the perpendicular direction of
the Galactic plane(dashed line).
The circle indicates a tidal radius of NGC 6626.
}
\label{N6626contour} 
\end{figure}

The top-left to bottom-right panels of Figure~\ref{N6626contour} show the star count map of NGC 6626,  the isodensity contour map smoothed by 
a Gaussian kernel  value of  $0^{\circ}.045$ and $0^{\circ}.12$, and the distribution of E(B-V) extinction value of ~\citet{Sch98}.
The isodensity contour lines correspond to the level of 
$1.0\sigma, 1.5\sigma, 2.0\sigma, 3.0\sigma, 4.0\sigma$, and $6.0\sigma$.
The proper motion of $\mu_{\alpha}\cos\delta=0.63\pm0.67$ mas yr$^{-1}$ and  $\mu_{\delta}=-8.46\pm0.67$
mas yr$^{-1}$~\citep{Casetti2013} is represented by a long arrow.
The dashed line and solid line indicate the direction of the Galactic center and the perpendicular direction of the Galactic plane, respectively.
The tidal radius of NGC 6626 (i.e., $r_t=11'.27$) from~\citet{Har96} is also plotted as a circle.

In Figure~\ref{N6626contour}, it is apparent that the stellar density distribution around NGC 6626
shows distorted overdensity features and extended tidal tails beyond the tidal radius. 
The tidal tails seem to stretch out symmetrically to both
sides of the cluster, extending toward east and west directions from the cluster center to the radial distance of $\sim2r_t$.
In addition, the two tidal tails are likely to be aligned with the directions of the Galactic center and anti-center.
Although there is no apparent extending features toward the direction of the
proper motion in the surface density maps, there is a clumpy substructure in the northern area, which is
aligned with the opposite direction of proper motion.
~\citet{Chun2012} first found the prominent overdensity feature that extends toward the
perpendicular direction to the Galactic plane within the tidal radius of NGC 6626. 
We also found a stellar substructure similar to that found by~\citet{Chun2012} in the contour map with a kernel 
values of $0^{\circ}.045$.
This substructure extends toward north-west direction within the tidal radius but is not as prominent as the substructure observed 
by~\citet{Chun2012}.
We note here that our spatial density distribution has a wider field of view than that of ~\citet{Chun2012}, which has enabled us
to estimate and calibrate the underlying background substructure more accurately. Thus, the stellar density structure in this
study is more homogeneous and less affected by field star contamination.

\begin{figure}
\centering
\includegraphics[width=0.47\textwidth]{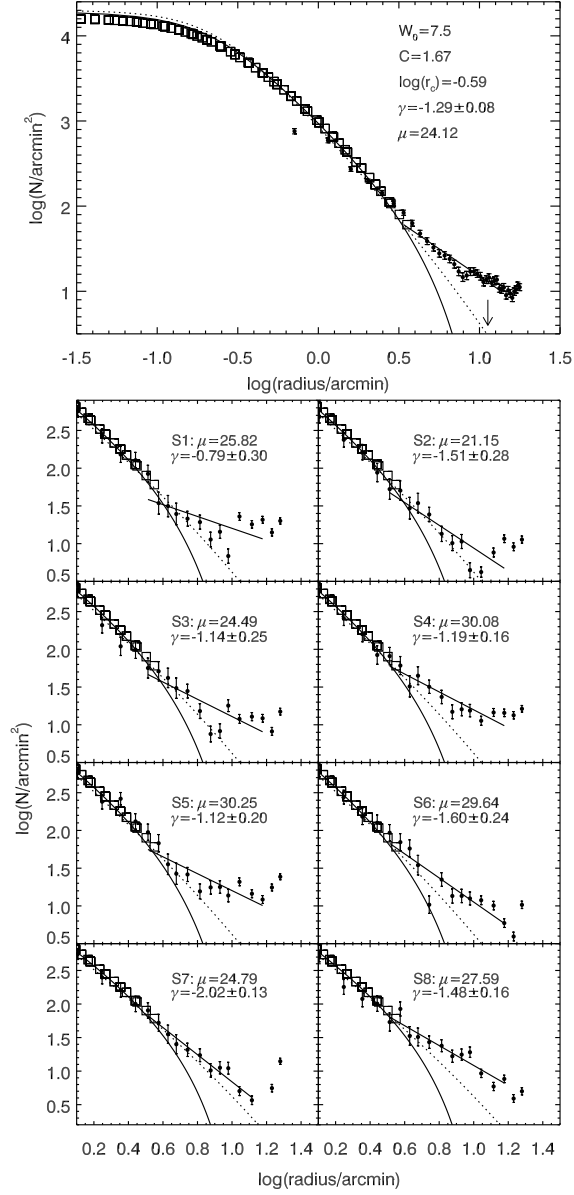}
\caption{{\it Upper} : Radial surface density profile of NGC 6626 with a King
model (solid curve) and a Wilson model (dotted curve).
The profile within the detected overdensity region, $\sim 0.28 r_t\leq r\leq \sim 1.5 r_t$,
is represented by a power-law, i.e., a dotted straight line in logarithmic scale
with a slope of $\gamma=-1.29\pm0.08$.
The mean number density of stars in the overdensity region is estimated
as $\mu=24.12$ per arcminute square.
{\it Lower} : Radial surface density profiles in eight different angular
sections (S1$\sim$S8). For clarity, we magnified the radial profile of the overdensity feature.
The other notations are the same as those of the
radial surface density profile in the upper panel.
}  
\label{N6626radial} 
\end{figure}

A radial surface density profile for NGC 6626 was presented in the upper panel of Figure~\ref{N6626radial}.
In the central region of the cluster, the number density profile was substituted with the surface brightness profile of~\citet{Tra95},
which connects to the number density profile at the middle range. The theoretical King model and Wilson model were also plotted
to characterize the observed radial profile. It is apparent that  our number density profile does not trace the
King model and Wilson model at the outer region of the cluster; 
instead, it shows overdensity feature with a break in the slope of profile at the radius of log($r'$)$\sim0.5$
$(\sim0.28 r_t)$. 
The overdensity feature extends out to the log($r'$)$\sim1.1$ $(\sim1.5 r_t)$, 
and the profile in this region is characterized by a power law with a slope of $\gamma=-1.29\pm0.08$. 
This slope is not very different from the slope of $\gamma = -1$, predicted for a constant mass-loss rate over a long time ~\citep{Joh99}.
The overdensity feature in the radial profile is indicative of extended tidal tails and substructures shown
in Figure~\ref{N6626contour}.

The radial surface density profiles of eight angular sections for NGC 6626 are plotted in
the lower panel of Figure~\ref{N6626radial}.
In general, all the radial profile show the overdensity features with a break in slope at the outer region of the cluster.
The estimated mean surface densities $(\mu)$ in angular sections 4,5 and 6 are higher than those in the other sections,
and the slopes ${\gamma}$ of profiles in section 4 and 5 are somewhat shallower than those in the other sections.
Furthermore, the radial profiles in section 4 and 5 still maintain the overdensity features at the radial distance of  log($r'$)$\sim1.3$.
These overdensity features correspond to the apparent extratidal tails extending toward the direction perpendicular to the 
Galactic plane and direction of the Galactic center, as shown in Figure~\ref{N6626contour}.
Although the density in section 1 is not as high as that of section 5, this is because that
the density near the tidal radius is low.  
Indeed, the weak connection between the cluster and tails is shown in contour map with a low kernal value of $0^{\circ}.045$. 
However, the tail extends out to a radial distance of $2r_t$, and this density feature is represented by a shallow slope in the 
and high density at the outer radius in radial profile of section 1.

\subsection{NGC 6642}
\begin{figure}
\centering
\includegraphics[width=0.5\textwidth]{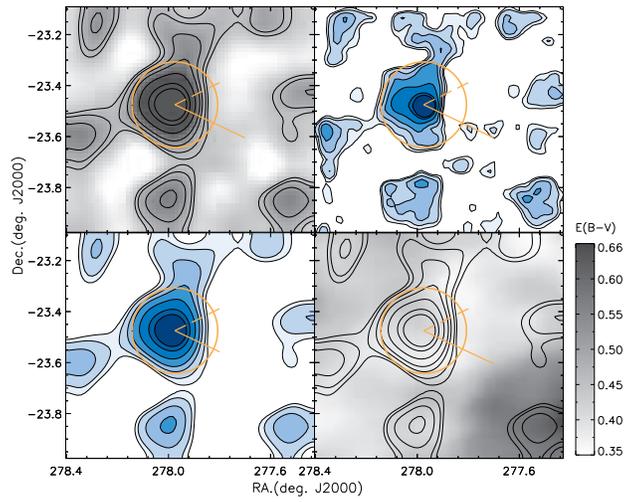}
\caption{From top-left to bottom-right, the star count map around NGC 6642, the surface
density maps smoothed by different Gaussian kernel values, and the E(B-V) extinction map around
the cluster.
The Gaussian kernel values are $0^{\circ}.045$ and $0^{\circ}.12$.
Isodensity contour levels are $0.5\sigma, 1.0\sigma, 2.0\sigma, 3.0\sigma, 5.0\sigma, 8.0\sigma$, and $10.0\sigma$.
The circle indicates a tidal radius of NGC 6642. 
Solid line indicates the direction of the Galactic center, and dashed line shows the perpendicular direction of
the Galactic plane.}
\label{N6642contour} 
\end{figure}

We plot star count map, isodensity contour maps, and the distribution map of 
E(B-V) value~\citep{Sch98} from the top-left to bottom-right panel in Figure~\ref{N6642contour}
in order to investigate of stellar distribution of NGC 6642. Gaussian kernel widths of $0^{\circ}.045$, and $0^{\circ}.12$
were applied to find spatial coherence in the stellar distribution. 
The isodensity contour levels are $0.5\sigma, 1.0\sigma, 2.0\sigma, 3.0\sigma, 5.0\sigma, 8.0\sigma$,
and $10.0\sigma$.
The contour lines in the star count map and the distribution map of E(B-V) correspond to those of a smoothed map
with a Gaussian kernel width of $0^{\circ}.12$.
The circle centered on the cluster indicates the tidal radius of $r_t=10.07^{'}$~\citep{Har96}.
The solid and dashed lines represent the direction of the Galactic center and a perpendicular direction to the 
Galactic plane, respectively.

As can be seen in Figure~\ref{N6642contour}, 
the stellar distribution of NGC 6642 seems to show a drop in density around the tidal radius in the specific direction, and
clumpy structures outside of tidal radius. 
However, the prominent extended stellar substructure elongates in a northern direction beyond the tidal radius. 
A clumpy chunk in the southern region seems to be a counterpart to the overdensity feature in the northern region. 
A marginal extension also appears in an eastern direction, which seems to be aligned with the opposite perpendicular 
direction to the Galactic plane. Unfortunately, we could not find substructures that might be associated with
the proper motion of the cluster, because the proper motion of NGC 6642 has not yet been reported.

\begin{figure}
\centering
\includegraphics[width=0.47\textwidth]{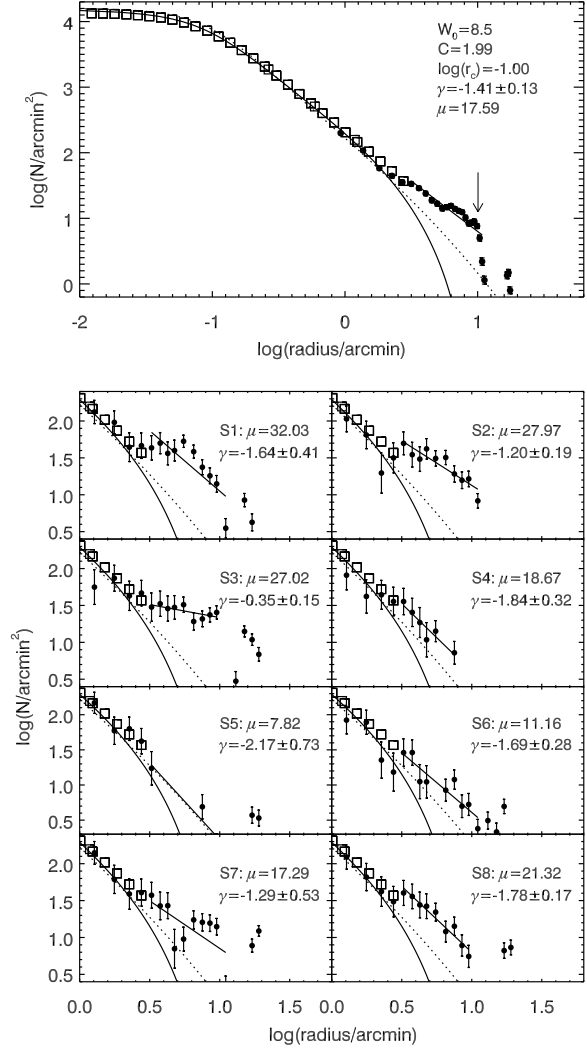}
\caption{{\it Upper} : Radial surface density profile of NGC 6642 with a King
model (solid curve) and a Wilson model (dotted line)
There is overdensity feature at $\sim 0.3 r_t\leq r\leq \sim 1.5 r_t$, which
is represented by a power-law, i.e., a dotted straight line in logarithmic scale
with a slope of $\gamma=-1.41\pm0.13$.
The mean number density of stars in the overdensity region is estimated
as $\mu=17.59$ per arcminute square.
{\it Lower} : Radial surface density profiles of eight different angular
sections (S1$\sim$S8).
The radial density points, which were calculated by interpolation from neighbor densities, were indicated
by open triangle.
The other notations are the same as those of the
radial surface density profile in the upper panel.
}  
\label{N6642radial} 
\end{figure}

The upper panel of Figure~\ref{N6642radial} shows the radial surface density
profile of NGC 6642 along with the King model, the Wilson model and surface brightness profile of ~\citet{Tra95}.
The radial surface profile of NGC 6642 shows apparent overdensity feature that depart from both model predictions
from the radius of log($r'$)$\sim0.5$ $(\sim0.3r_t)$ to the tidal radius. However, the overdensity feature does not continue, and
the density of radial profile decreases abruptly at the tidal radius. 
The drop in density and local clumpy substructures around/outside tidal radius shown in Figure~\ref{N6642contour}
seem to be associated with this fall of the radial profile and low density feature at the outer region.
The overdensity feature within tidal radius was fitted by a power law with a slope $\gamma=-1.27\pm0.10$.


The radial surface density profiles for eight angular sections, which are plotted in the lower
panel of  Figure~\ref{N6642radial}, represent better the stellar density distribution around
NGC 6642 than average radial density profile in the upper panel of Figure~\ref{N6642radial}.
The radial profiles in angular sections 1 and 8 show clear overdensity features within tidal radius, and section 1 has the largest mean 
density of the eight sections. Although some radial points in these sections are not plotted because of small number statistics, 
there are still considerable number densities at the outer region.
This is in agreement with the overdensity feature extending toward the eastern side on the two-dimensional surface density map.
In contrast to these angular sections, the overdensity feature in section 2 is disconnected at the tidal radius because of its
low density at the outer region. 
In the two dimensional surface contour map, we can not find obvious stellar substructures in that region.
The radial density profile in section 3 shows the most prominent overdensity features with high density and the flattest slope, and contains
the density excess at the outer radius. These are representative of a prominent extending substructure in northern region 
on the isodensity contour map in Figure~\ref{N6642contour}.
In the section 7, the radial density profile is likely to show the overdensity feature associated with the counterpart of the 
extended substructure in the northern region.
In the sections 4 and 5, we could not find clear overdensity feature, and
two-dimensional contour map does not also show prominent stellar substructures in those regions.


\subsection{NGC 6723}
\begin{figure}
\centering
\includegraphics[width=0.5\textwidth]{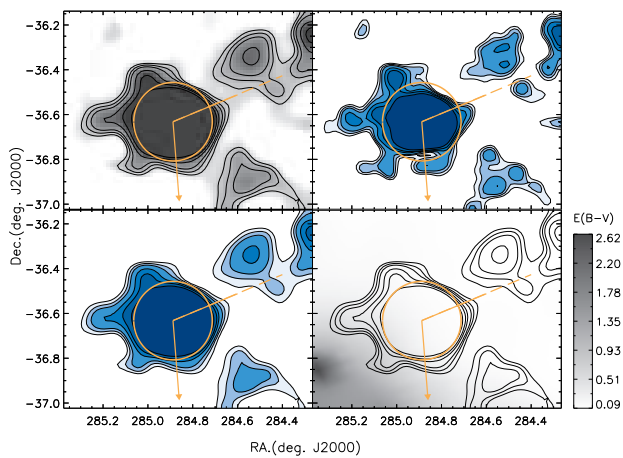}
\caption{From top-left to bottom-right panels, the star count map around NGC 6723, the surface
density maps smoothed by Gaussian kernel values of $0^{\circ}.07$ and $0^{\circ}.11$, the distribution map
of E(B-V). 
Isodensity contour levels are $2.0\sigma, 2.5\sigma, 3.0\sigma, 4.0\sigma, 5.0\sigma$, and $7.0\sigma$.
The circle indicates a tidal radius of NGC 6723. The arrow represent the proper motion of cluster.
Solid line indicates the direction of the Galactic center, and dashed line shows the perpendicular direction of
the Galactic plane. We note that the direction to the Galactic center and the Galactic plane almost coincide, thus
two lines overlap each other.}
\label{N6723contour} 
\end{figure}

Figure~\ref{N6723contour} shows a star count map around NGC 6723,
surface density maps smoothed with Gaussian kernel values
of $0^{\circ}.07$ and $0^{\circ}.11$ and a distribution map of E(B-V) value~\citep{Sch98}
from the upper-left panel to the lower-right panel. 
We note that different Gaussian kernel values for NGC 6723 were selected in order to
highlight the structures with similar spatial extents.
Isodensity contours were overlaid on the maps with contour levels of
$2.0\sigma, 2.5\sigma, 3.0\sigma, 4.0\sigma, 5.0\sigma$, and $7.0\sigma$.
The long dashed line and solid line
indicate the perpendicular direction to the Galactic plane and the direction of the Galactic center, respectively.
The proper motion of NGC 6723, i.e., $\mu_{\alpha}\cos\delta=-0.17\pm0.45$ mas yr$^{-1}$ and $\mu_{\delta}=-2.16\pm0.50$ mas
yr$^{-1}$ ~\citep{Din2003}, was indicated with an arrow.
The tidal radius of $r_t=10'.51$~\citep{Har96} was represented by circle.

As can be seen in Figure~\ref{N6723contour}, there are weak extended substructures beyond the tidal radius of NGC 6723 at 
levels larger than $\sigma$.
Small density lobes are likely to extend toward the eastern and western sides;
the direction of the Galactic center, perpendicular direction to the Galactic plane or their opposite directions.
In addition, the isodensity contour lines show a horn-shaped structure in the northern region, which corresponds to the opposite direction
of the proper motion.
The marginal extension appears near the tidal radius in the southern region, 
but this weak substructure does not seem to be outgrowing anymore.
We note that there is a huge reflection nebula in the distant souther region of the cluster.
Thus dust in the outskirts of reflection nebula in the southern region could affect the detected number
density of marginal extension in the southern region. Indeed, the value of $E(B-V)$ in the southeast region is higher than
the values of other regions. Thus, there is a possibility that more extended stellar substructures could exist in the obscured region.

\begin{figure}
\centering
\includegraphics[width=0.47\textwidth]{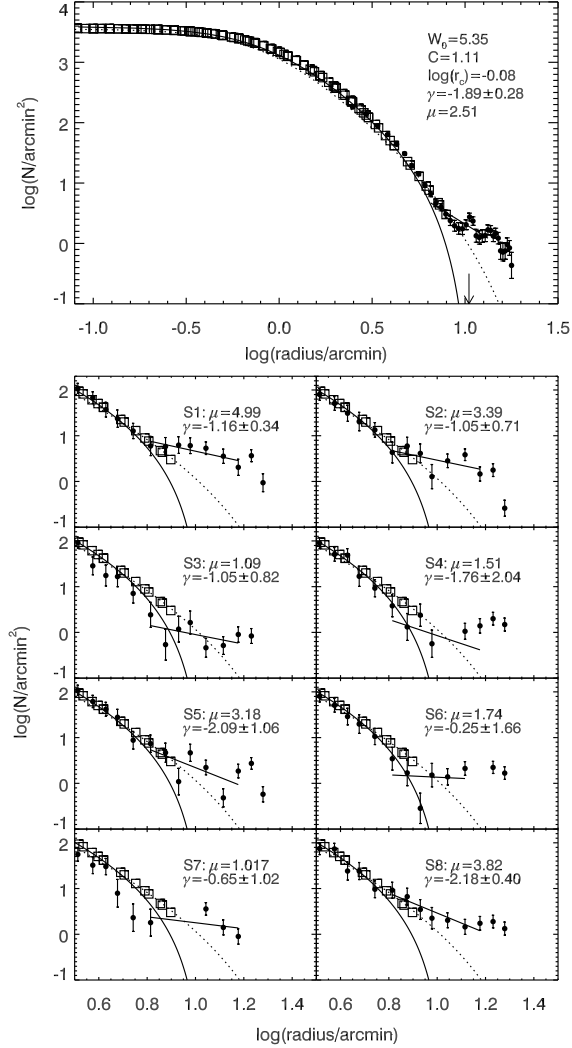}
\caption{{\it Upper} : Radial surface density profile of NGC 6723 with a King
model (solid line) and a Wilson model (dotted line).
The overdensity feature appears at the region of $\sim 0.6 r_t\leq r\leq \sim 1.5 r_t$.
The overdensity feature are represented by a power-law, i.e., a dotted straight line in logarithmic scale
with a slope of $\gamma=-1.89\pm0.28$.
The mean number density of stars in the overdensity region is estimated
as $\mu=2.51$ per arcminute square.
{\it Lower} : Radial surface density profiles of eight different angular
sections (S1$\sim$S8).
For clarity, we magnified the radial profile of the overdensity feature.
The other notations are the same as those of the
radial surface density profile in the upper panel.
}  
\label{N6723radial} 
\end{figure}

The radial surface densities for NGC 6723, measured in each concentric annulus, are shown
in the upper panel of Figure ~\ref{N6723radial}.
The theoretical King model and Wilson model were arbitrarily normalized to our measurements, and
the surface brightness profile of~\citet{Tra95} was used as a substitute for our measurements in the central region.
Apparently, the radial number density profile departs from the theoretical King model at the outer region of the cluster, while
Wilson model has a slightly better fit to our radial profile of the cluster.
The overdensity feature, which departs from the King model with a break in slope, is shown at the radius of log($r'$)$\sim$0.8 ($\sim$0.6$r_t$),
and extends to the radius of log($r'$)$\sim$1.15 ($\sim$1.5$r_t$). 
The slope in this overdensity region is characterized by a power law with a slope of $\gamma =-1.89\pm0.28$, which is steeper than 
the value predicted from a theoretical simulation with a constant orbit-averaged mass-loss rate~\citep{Joh99}.

The radial surface profiles of eight angular sections are also shown in the lower panel of Figure~\ref{N6723radial}.
The overdensity feature in the region of $0.6r_t\lesssim r \lesssim 1.5r_t$
is commonly detected in specific angular sections.
The radial profile in section 1, where the prominent substructure appears in the two-dimensional contour map, has the highest 
mean number density out of all the profiles of the eight sections. In addition, the densities in section 2, 5, and 8 have somewhat 
higher density levels in the overdensity region, which is in good agreement with the horn-shaped substructure, the side lobe in the west side, 
and the marginal structure in the south region shown in the Figure~\ref{N6723contour}.
Some radial points near tidal radius in section 7 were not plotted, because of its low number density. 
This low density feature also appears as a bay-shaped structure with low level contours in the southern region on the contour map with
a kernel value of $0^{\circ}.07$.
The possible narrow dust lane, which might extend from the reflection nebular in southern region, can cause this low number density feature.
Unfortunately, the extinction map of~\citep{Sch98} does not show an apparent dust lane because of its low resolution.
Thus, the study of dust extinction with high resolution is necessary in order to study stellar distribution around the cluster.

\section{DISCUSSION}
All of our target clusters reside within $\sim3$ kpc from the Galactic center. 
The ancient globular clusters in the inner region can provide clues regarding the formation of the Galactic bulge and disk.
Indeed, the spatial, chemical and kinematical properties of several globular clusters in the inner regions 
are consistent with those of bulge, disk and even bar membership~\citep{Burkert1997,Barbuy1998,Cote1999,Heitsch1999}.
However, the definite decomposition of globular clusters between the bulge and disk or bar components is not a
trivial task in the central region of the Galaxy because the inner regions of the Galaxy are superimposed by 
various stellar populations of bulge, disk, and bar. 
In addition, in the inner region of the Galaxy, the globular clusters experience extreme dynamical evolution due to 
bulge/disk shock and strong tidal effects~\citep{Aguilar1988,Shin2008}.
Thus, reliable measurements of metallicities, orbits, and distances for globular clusters in the central regions are necessary
in order to understand the evolution of the globular cluster in the bulge region.
In this section, we investigated the previous results of our target globular clusters, and discuss the properties of clusters that 
could affect the substructures around the globular clusters.

NGC 6266 is a high-density~\citep[log$\rho_c\sim5.34$;][]{Jacoby2002,Possenti2003} and the ninth most luminous globular 
cluster located $\sim1.7$ kpc from the Galactic center~\citep{Har96}.~\citet{Din2003} found that NGC 6266 has a large
rotation velocity, suggesting that this cluster belongs to a disk system rather than to a pressure-supported system.
The total destruction rate of this cluster is about $\nu_{tot}=0.644$ per Hubble time, and destruction rate ratio is
$\nu_{tot}/\nu_{evap}=1.0$~\citep{Gnedin1997}, where $\nu_{evap}$ is evaporation rate per Hubble time. These destruction
rate and ratio indicate that the main process of dynamical evolution for this cluster is internally driven, such as two-body relaxation.
In this study, we found that apparent marginal extensions around the cluster bent to the direction of
proper motion and were likely to show S-shaped features. Thus, we can interpret this
orientation of stellar substructure as the signature that the evaporated stars by internal two-body relaxation are finding
themselves in a Galactic orbit similar to that of the parent cluster. However, we also found that
the stellar density configuration around NGC 6266 shows a spatial coherence associated with the effect of
dynamical interaction with the Galaxy. The prominent density feature extended toward the direction of
the Galactic center.
Thus, our stellar density distribution around NGC 6266 could be interpreted as an example of dynamical cluster evolution that
tidal shocking accelerates two-body relaxation~\citep{Kundic1995,Din99}.
~\citet{Lee07} classified NGC 6266 as an extended blue horizontal branch (EHB) cluster, and also suggested 
that the clusters with EHB could be a remnant of the first building blocks in the early universe. Thus, the stellar density feature
around NGC 6266 would be tidally associated with these unknown first building blocks. 
More accurate orbit information and an investigation of the field stars contamination around NGC 6266 are necessary to understand 
the dynamical evolution of NGC 6266.

NGC 6626 is massive and moderately metal-poor globular cluster in the Galactic bulge region.
~\citet{Din99} first found the thick-disk orbit for this cluster and then proposed the possibility that
this cluster had been produced in a satellite galaxy and then departed from its parent during the accretion process.
In addition, according to~\citet{Lee07}, this cluster is
moderate EHB cluster and could be a remnant of first building blocks. 
~\citet{Chun2012} first found the stellar density substructure around NGC 6626, which extends toward the direction
perpendicular to the Galactic plane, indicating a disk-shock effect. They also discussed the
possibility of the accretion scenario for the origin of this cluster. 
More recently,~\citet{Casetti2013} updated the proper motion and orbit of this cluster. They 
noted that the cluster is located at the apocenter and that its orbit is rather eccentric and disruptive, indicating
that the cluster may experience substantial mass loss. In our study, we did find prominent stellar substructures
extending toward the Galactic center and anti-center directions. This spatial orientation of the tidal tails is in good agreement with
the findings of ~\citet{Mon07}; namely, that the inner tails are oriented toward the Galactic center direction in the apocenter position.
Moreover, we confirmed the stellar substructure extending toward the perpendicular direction to the Galactic plane, which was
found by ~\citet{Chun2012}.~\citet{Casetti2013} mentioned that their newly updated proper motion vector with the Solar
motion subtracted is aligned with this substructure. They then speculated that the recent disk plane crossing about 4 Myr ago
might have contributed to the construction of tidal extension in the perpendicular direction to the Galactic plane. 
Therefore, this cluster is very likely to suffer from the disk/bulge shock and the Galactic tidal force. 
However, the total destruction rate of $\nu_{tot}=0.546$ per Hubble time and destruction ratio of
$\nu_{tot}/\nu_{evap}\sim1.0$ by ~\citet{Gnedin1997} do not seem to be in agreement with our interpretation.
Thus, we here note that recalculation of the destruction rate is necessary using recently updated proper motion and eccentric orbit.

NGC 6642 has dense central core with a high concentration $c=1.99$~\citep{Har96}. Previous studies~\citep{Tra95,Balbinot2009}
have also indicated that there is not a well-resolved core in the radial profile and have classified the cluster as core-collapsed cluster.
~\citet{Barbuy2006} noted that the age of NGC 6642 is comparable to that of M5 and then suggested that the cluster
is one of the few genuine metal-poor and old clusters in the bulge region. Therefore it can contain fossil information about the Galaxy.
On the other hand, ~\citet{Balbinot2009} examined the position of NGC 6642 in the HB type versus the metallicity diagram
and found that NGC 6642 lies on the position of young halo populations, not on the position of
old halo and disc/bulge populations. They concluded that NGC 6626 is a transition 
cluster between the inner halo and outer bulge, considering its age, its position in the Galaxy, and its HB morphology.
They also found clear evidence for the mass segregation
and depletion of low-luminosity stars in the luminosity and mass function and attributed this
dynamical structure to the disk and bulge shocking.
Indeed, we did find clumpy stellar substructures which extend toward north direction and the opposite
direction perpendicular to the Galactic plane beyond the tidal radius. 
The large total destruction rate of $\nu_{tot}=1.90$ per Hubble time 
and destruction ratio of $\nu_{tot}/\nu_{evap}\sim8.3$ by ~\citet{Gnedin1997} also indicate 
that a strong environmental force, such as a tidal effect due to disk and bulge shocking,
might affect the dynamical evolution of this cluster.
Thus, we interpret that although NGC 6642 is a core-collapse cluster and two-body relaxation would affect
the dynamical evolution of the cluster, the extended stellar substructures around NGC 6642 
would be the result of a strong gravitational interaction with the Galaxy.
Unfortunately, we were unable to
investigate a definite association between the observed stellar substructure and the cluster's motion because
there is no published proper motion information for this cluster. If we could obtain accurate proper motion for this cluster,
we would be able to understand the dynamical evolution of the cluster and figure out the origin of the cluster; e.g., whether
this cluster is a genuine metal-poor and old cluster in the bulge region or a transition cluster between the inner halo region and
the outer bulge region.

NGC 6723, which has a relatively low concentration of density~\citep[$c=1.05$,][]{Har96} and a moderate EHB morphology~\citep{Lee07},
is an old globular clusters in the Galactic bulge region.
Although it has long been considered one of the genuine Galactic bulge globular clusters, the origin of the cluster
is not obvious because there have only been few studies on this cluster.
~\citet{Van1993} suggested that NGC 6723 would have a circular orbit motion in the central region of the Galaxy. However,
~\citet{Din2003} measured the proper motion of the cluster and found that its orbit is highly inclined. 
Based on its kinematics and low metallicity, they concluded
that NGC 6723 is a member of the halo system. 
Recently,~\citet{Lee07} suggested a different origin scenario that metal-poor EHB globular clusters such as NGC 6723 
would be relics of the first building blocks.
The total destruction rate of $\nu_{tot}=0.321$ per Hubble time and 
a rather high destruction ratio of $\nu_{tot}/\nu_{evap}\sim2.0$ by~\citet{Gnedin1997} indicate that there would be weak tidal
substructures around this cluster.
Indeed, in our study we have found the tidal stripping features that are likely to be associated with the
interaction with the Galaxy.
The weak stellar density lobes, which extend toward the directions of the Galactic center and anti-center,
were detected in the surface iso-density map.
In addition, a marginal extension tracing the opposite direction of the proper motion was also detected.
However, we could not exclude dust extinction effect on the stellar density distribution map.
The thin dust lane of reflection nebula in the southern region of the cluster might hide the stellar extensions. 

\section{Summary}
In this study, we investigated the stellar spatial density distribution around four metal-poor globular clusters
(NGC 6266, NGC 6626, NGC 6642, and NGC 6723) in the Galactic
bulge region using wide-field ($45'\times45'$) near-infrared
$J, H,$ and $K$ images obtained with the WFCAM camera on the UKIRT. In order to discard the field stars contamination and
enhance density contrast between cluster member and field stars, we used the C-M mask filtering technique and
the optimal contrast filtering technique on the cluster CMDs. Two-dimensional density contour maps of the clusters were
examined, and radial density profiles were also investigated with King and Wilson models.

The two-dimensional stellar density contour maps for the four globular clusters showed asymmetric and extended features of the 
stars around the globular clusters. In particularly, three globular clusters (NGC 6266, 
NGC 6626, and NGC 6642) showed tidal extension and a stellar chunk beyond the tidal radius, while NGC 6723 showed weak density lobes 
near the tidal radius.  Such extended stellar substructures were aligned with the direction of the Galactic center and anti-center, the 
perpendicular direction to the Galactic plane or the direction of the cluster orbit motion. 
Thus, it is highly probably that target clusters are affected by the dynamical environment effect such as a tidal force and bulge/disk shock.
The radial profile also represented the extended substructures in the two-dimensional contour maps as overdensity features
with break in a slope.
Although the observed radial profiles of clusters departed from the King and Wilson model at the outer region of cluster,
Wilson model appeared to have a better fit with the observed radial density profile for the cluster, which seemed to experience less 
of an dynamical environment effect (e.g. NGC 6266 and NGC 6723).

We here note that internal driven such as two-body relaxations, also affects the extended stellar substructure around the globular clusters.
The stellar extensions around NGC 6266 show a good example of internal driven.
In this cluster, the stars might be primarily evaporated by two-body relaxation, and then be affected by environmental tidal force
of the Galaxy. For other clusters, it is obvious that the two-body relaxation contributes to the extended stellar substructures around clusters.
In contrast to the case of NGC 6266, however, the environmental effects such as tidal force of the Galaxy and bulge/disk
shock would be main factors for the extended stellar substructures, even though the cluster such as NGC 6642 had very active two-body relaxations.
In addition, a more accurate proper motion and orbit that is calculated using an accurate galaxy potential model 
could increase the portion of external effect on the dynamical evolution of the globular clusters in the bulge region.
Indeed, although the tidal-shock rate of NGC 6626 was not comparable or larger
than two-body relaxation rates in previous studies~\citep{Gnedin1997,Din99}, 
the orbit of NGC 6626, which was derived using more accurate proper motion
and an axisymmetric and barred model of the Galaxy~\citep{Casetti2013}, was found to be eccentric and more disruptive. 
Therefore, further studies of the dynamical evolution of the globular clusters in the bulge region with accurate proper motion and
a galaxy model are required to provide theoretical constraints of the configuration of stars around the globular clusters.
In addition, further deep and wide-field photometric data for metal-poor and metal-rich globular clusters in the Galactic
bulge could provide more accurate information about the dynamical evolution of globular clusters, thereby leading to an increased 
understanding of the origin of globular clusters in the bulge region, as well as the formation of the bulge region.

\acknowledgments
We are grateful to an anonymous referee for detailed comments that greatly improved this paper.
This research was supported by the Basic Science Research Program through
the National Research Foundation of Korea (NRF) funded by the Ministry of
Education, Science and Technology (2013R1A1A2006826). 
This work is grateful for partial support from KASI-Yonsei DRC program of Korea Research Council of
Fundamental Science and Technology (DRC-12-2-KASI).

%

%

%

%

%

\clearpage

%

\clearpage

%

\clearpage

%

\begin{deluxetable}{ccccccccc}
\tabletypesize{\scriptsize}
\tablewidth{0pt}
\tablecaption{Basic parameter for four target globular clusters and position of three comparison fields\label{para}}
\tablehead{
\colhead{Target} & \colhead{$\alpha$} & \colhead{$\delta$} &
\colhead{$R_{sun}$} & \colhead{$R_{GC}$} & \colhead{$r_c$} & 
\colhead{$r_t$} & \colhead{[Fe/H]} & \colhead{Index}\\
 & \colhead{(J2000)} & \colhead{(J2000)} &
 \colhead{(kpc)} & \colhead{(kpc)} & \colhead{($'$)} & \colhead{($'$)} &
}
\startdata
NGC 6266 & 17:01:12.80 & -30:06:49.4 & 6.8 & 1.7 & 0.22 & 8.97 & --1.18 & -\\
NGC 6626 & 18:24:32.81 & -24:52:11.2 & 5.5 & 2.7 & 0.24 & 11.27 & --1.32 & -\\
NGC 6642 & 18:31:54.10 & -23:28:30.7 & 8.1 & 1.7 & 0.1 & 10.07 & --1.26 & -\\
NGC 6723 & 18:59:33.15 & -36:37:56.1 & 8.7 & 2.6& 0.83 & 10.51 & --1.10 & -\\
Comparison1 & 18:31:37.44 & -29:19:30.36 & - & - & - & - & - & NGC 6266, NGC 6642\\
Comparison2 & 17:12:59.28 & -23:12:51.84 & - & - & - & - & - & NGC 6626\\
Comparison3 & 19:09:22.32 & -32:55:42.96 & - & - & - & - & - & NGC 6723
\enddata
\tablecomments{ 
$R_{sun}$ and $R_{GC}$ are distances from the Sun and the Galactic center, 
respectively. $r_c$ and $r_t$ indicate the core radius and tidal radius.
The basic parameter information is from the catalogue~\citealt{Har96} (2010 edition). Index indicates
the globular clusters to which comparison fields applied in C-M mask filtering and optimal contrast filtering technique.
}
\end{deluxetable}

\begin{deluxetable}{ccccc}
\tabletypesize{\scriptsize}
\tablewidth{0pt}
\tablecaption{Observation Summary of four globular clusters\label{log}}   
\tablehead{
\colhead{Target} & \colhead{Filter} &  \colhead{Exp. time} &\colhead{FWHM} \\
& & \colhead{(micro-step $\times$ dither $\times$ second)} & \colhead{$(^{''})$}
}    
\startdata
NGC 6266 & $J$     & $4\times5\times1s, 4\times5\times5s$ &  0.78, 0.77 \\      
         & $H$     & $4\times5\times1s, 4\times5\times5s$ &  0.78, 0.76 \\
         & $K$ & $4\times5\times1s, 4\times5\times10s$ &  0.74, 0.77 \\
NGC 6626 & $J$     & $4\times5\times1s, 4\times5\times5s$ &  0.92, 0.90 \\      
         & $H$     & $4\times5\times1s, 4\times5\times5s$ &  0.88, 1.05 \\
         & $K$ & $4\times5\times1s, 4\times5\times10s$ &  0.91, 0.76 \\
NGC 6642 & $J$     & $4\times5\times1s, 4\times5\times5s$ &  0.86, 0.87 \\      
         & $H$     & $4\times5\times1s, 4\times5\times5s$ &  0.81, 0.79 \\
         & $K$ & $4\times5\times1s, 4\times5\times10s$ &  0.82, 0.78 \\
NGC 6723 & $J$     & $4\times5\times1s, 4\times5\times5s$ & 0.95, 0.99 \\      
         & $H$     & $4\times5\times1s, 4\times5\times5s$ & 1.07, 1.06 \\
         & $K$ & $4\times5\times1s, 4\times5\times10s$ & 0.98, 0.97                
\enddata                               
\end{deluxetable}

\begin{deluxetable}{ccc}
\tabletypesize{\scriptsize}
\tablewidth{0pt}
\tablecaption{The coefficient $a$ and $b$ of new color indices for each cluster\label{coefficients}}   
\tablehead{
\colhead{Target} & \colhead{a} & \colhead{b} 
}    
\startdata
NGC 6266 &  $0.748$  &  $0.663$  \\      
NGC 6626 &  $0.784$  &  $0.621$  \\      
NGC 6642 &  $0.757$  &  $0.653$  \\      
NGC 6723 &  $0.750$  &  $0.661$    
\enddata                               
\end{deluxetable}
\clearpage


\begin{thebibliography}{}
\bibitem[Abadi et al.(2006)]{Aba06}Abadi, M. G., Navarro, J. F., \& Steinmetz, M.
    2006, \mnras, 365, 747
\bibitem[Aguilar et al.(1988)]{Aguilar1988} Aguilar, L., Hut, P., 
\& Ostriker, J.~P.\ 1988, \apj, 335, 720    
\bibitem[Balbinot et al.(2009)]{Balbinot2009} Balbinot, E., 
Santiago, B.~X., Bica, E., \& Bonatto, C.\ 2009, \mnras, 396, 1596
\bibitem[Baugh et al.(1996)]{Bau96}Baugh, C. M., Cole, S., \& Frenk, C. S. 1996, \mnras, 283, 1361
\bibitem[Barbuy et al.(1998)]{Barbuy1998} Barbuy, B., Bica, E., \& Ortolani, S.\ 1998, \aap, 333, 117
\bibitem[Barbuy et al.(2006)]{Barbuy2006} Barbuy, B., Bica, E., Ortolani, S., \& Bonatto, C.\ 2006, \aap, 449, 1019
\bibitem[Bellazzini et al.(2003)]{Bel03}Bellazzini, M., Ferraro, F. R., \& Ibata, R. 2003,
    \aj, 125, 188
\bibitem[Belokurov et al.(2006a)]{Bel06a}Belokurov, V., Zucker, D. B., Evans, N. W., et al. 2006a, \apj, 642, L137
\bibitem[Belokurov et al.(2006b)]{Bel06b}Belokurov, V., Evans, N. W., Irwin, M. J., Hewett P. C., \&
    Wilkinson, M. I. 2006b, \apj, 637, L29
\bibitem[Belokurov et al.(2007)]{Bel07}Belokurov, V., Evans, N. W., Irwin, M. J., et al. 2007, \apj, 658, 337
\bibitem[Bertin et al.(2002)]{Bertin2002}Bertin, E., Mellier, Y., Radovich, M., Missonnier, G.,
	 Didelon, P., \& Morin, B. 2002, in ASP Conf. Proc., Vol. 281, Astronomical Data Analysis Software
	and Systems XI, ed. D. A. Bohlender, D. Durand, \& T. H. Handley (San Francisco, CA: ASP), 228 
\bibitem[Bullock \& Johnston(2005)]{Bul05} Bullock, J. S., \& Johnston, K. V. 2005, \apj, 635, 931
\bibitem[Bullock et al.(2001)]{Bul01}Bullock, J. S., Kratsov, A. V., \& Weinberg, D. H.
    2001, \apj, 548, 33
\bibitem[Burkert \& Smith(1997)]{Burkert1997} Burkert, A., \& Smith, G.~H.\ 1997, \apjl, 474, L15
\bibitem[Casetti-Dinescu et al.(2013)]{Casetti2013} Casetti-Dinescu, D.~I., Girard, T.~M., J{\'{\i}}lkov{\'a}, L., et al.\ 
2013, \aj, 146, 33 
\bibitem[Carballo-Bello et al.(2012)]{Carballo2012} Carballo-Bello, 
J.~A., Gieles, M., Sollima, A., et al.\ 2012, \mnras, 419, 14
\bibitem[Carretta et al.(2009)]{Carretta2009} Carretta, E., Bragaglia, A., Gratton, R., D'Orazi, V., \& Lucatello, S.\ 2009, \aap, 508, 695
\bibitem[Chun et al.(2012)]{Chun2012} Chun, S.-H., Kim, J.-W., Kim, M.~J., et al.\ 2012, \aj, 144, 26
\bibitem[Chun et al.(2010)]{Chu10}Chun, S.-H., Kim, J.-W., Sohn, S. T., et al. 2010, \aj, 139, 606
\bibitem[Combes et al.(1999)]{Com99}Combes, F., Leon, S., \& Meylan, G. 1999, \aap, 352, 149
\bibitem[Conroy et al.(2011)]{Conroy2011} Conroy, C., Loeb, A., 
\& Spergel, D.~N.\ 2011, \apj, 741, 72
\bibitem[C{\^o}t{\'e}(1999)]{Cote1999} C{\^o}t{\'e}, P.\ 1999, \aj, 118, 406
\bibitem[Diemand et al.(2007)]{Diemand2007} Diemand, J., Kuhlen, 
M., \& Madau, P.\ 2007, \apj, 667, 859 
\bibitem[Dinescu et al.(2003)]{Din2003} Dinescu, D.~I., Girard, 
T.~M., van Altena, W.~F., \& L{\'o}pez, C.~E.\ 2003, \aj, 125, 1373 
\bibitem[Dinescu et al.(1999)]{Din99}Dinescu, D. I., Girard, T. M., \& van Altena, W. F.
    1999, \aj, 117, 1792
\bibitem[Drake et al.(2013)]{Drake2013} Drake, A.~J., Catelan, 
M., Djorgovski, S.~G., et al.\ 2013, \apj, 765, 154
\bibitem[Duffau et al.(2006)]{Duf06}Duffau, S., Zinn, R., Vivas, A. K., et al. 2006, \apj, 636, L97
\bibitem[Dye et al.(2006)]{Dye2006} Dye, S., Warren, S.~J., 
Hambly, N.~C., et al.\ 2006, \mnras, 372, 1227 
\bibitem[Ferraro et al.(2009)]{Fer09}Ferraro, F. R., Dalessandro, E., Mucciarelli, A., et al. 2009, \nat, 462, 483
\bibitem[Font et al.(2006)]{Fon06}Font, A. S., Johnston, K. V., Bullock, J. S., \& Robertson, B. E.
	2006, \apj, 638, 585
\bibitem[Gnedin \& Ostriker(1997)]{Gnedin1997} Gnedin, O.~Y., \& Ostriker, J.~P.\ 1997, \apj, 474, 223
\bibitem[Gratton et al.(2004)]{Gratton2004} Gratton, R., Sneden, C., \& Carretta, E.\ 2004, \araa, 42, 385
\bibitem[Grillmair et al.(2013)]{Grillmair2013} Grillmair, C.~J., 
Cutri, R., Masci, F.~J., et al.\ 2013, \apjl, 769, L23
\bibitem[Grillmair(2006)]{Gri06}Grillmair, C. J. 2006, \apj, 645, L37
\bibitem[Grillmair \& Dionatos(2006a)]{Gril06a}Grillmair, C. J., \& Dionatos, O. 2006a, \apj, 641, L37
\bibitem[Grillmair \& Dionatos(2006b)]{Gril06b} Grillmair, C. J., \& Dionatos, O.\ 2006b, \apjl, 643, L17
\bibitem[Grillmair et al.(1995)]{Gri95}Grillmair, C. J., Freeman, K. C., Irwin, M., \& Quinn, P. J.
    1995, \aj, 109, 2553
\bibitem[Grillmair \& Johnson(2006)]{Grill06}Grillmair, C. J., \& Johnson, R. 2006, \apj, 639, 17
\bibitem[Harris(1996)]{Har96}Harris, W. E. 1996, \aj, 112, 1487
\bibitem[Heitsch \& Richtler(1999)]{Heitsch1999} Heitsch, F., \& Richtler, T.\ 1999, \aap, 347, 455
\bibitem[Helmi et al.(1999)]{Hel99}Helmi, A., White, S. D. M., de Zeeuw, P. T., \& Zhao, H.
	1999, \nat, 402, 53
\bibitem[Ibata et al.(1994)]{Iba94}Ibata, R. A., Gilmore, G., \& Irwin, M. J. 1994, \nat, 370, 194
\bibitem[Ibata et al.(1995)]{Iba95}Ibata, R. A., Gilmore, G., \& Irwin, M. J. 1995, \mnras, 277, 781
\bibitem[Ibata et al.(2001)]{Iba01}Ibata, R., Irwin, M., Lewis, G., Ferguson, A. M. N., \&
     Tanvir, N. 2001, \nat, 412, 49
\bibitem[Ibata et al.(1997)]{Iba97}Ibata, R. A., Wyse, R. F. G., Gilmore, G., Irwin, M. J., \&
    Suntzeff, N. B. 1997, \aj, 113, 634
\bibitem[Ivezi\'{c} et al.(2000)]{Ive00}Ivezi\'{c} , \v{Z}., Goldston J., Finlator K., et al. 2000, \aj, 120, 963
\bibitem[Jacoby et al.(2002)]{Jacoby2002} Jacoby, B.~A., Chandler, A.~M., Backer, D.~C., Anderson, S.~B., 
\& Kulkarni, S.~R.\ 2002, \iaucirc, 7783, 1
\bibitem[Johnston(1998)]{Joh98}Johnston, K. V. 1998, \apj, 495, 297
\bibitem[Johnston et al.(2002)]{Joh02}Johnston, K. V., Choi, P. I., \& Guhathakurta, P.
    2002, \aj, 124, 127
\bibitem[Johnston et al.(1999)]{Joh99}Johnston, K. V., Sigurdsson, S., \& Hernquist, L.
    1999, \mnras, 302, 771
\bibitem[Juri\'{c} et al. (2008)]{Jur08}Juri\'{c}, M., Ivezi\'{c} , \v{Z}., Brooks, A., et al. 2008, \apj, 673, 864
\bibitem[Katz(1992)]{Kat92}Katz, N. 1992, \apj, 391, 502
\bibitem[King(1966)]{King66}King, I. R. 1966, \aj, 71, 64
\mnras, 400, 2162 
\bibitem[Klypin et al.(1999)]{Kly99}Klypin, A., Kravtsov, A. V., Valenzuela, O., \& Prada, F.
    1999, \apj, 522, 82
\bibitem[Koposov et al.(2012)]{Koposov2012} Koposov, S.~E., 
Belokurov, V., Evans, N.~W., et al.\ 2012, \apj, 750, 80
\bibitem[Koposov et al.(2010)]{Koposov2010} Koposov, S.~E., Rix, 
H.-W., \& Hogg, D.~W.\ 2010, \apj, 712, 260 
\bibitem[Kundic \& Ostriker(1995)]{Kundic1995} Kundic, T., \& Ostriker, J.~P.\ 1995, \apj, 438, 702
\bibitem[Law et al.(2009)]{Law2009} Law, D.~R., Majewski, 
S.~R., \& Johnston, K.~V.\ 2009, \apjl, 703, L67
\bibitem[Lee et al.(2009)]{Lee09}Lee, J.-W., Kang, Y.-W., Lee, J., \& Lee, Y.-W. 2009, \nat, 462, 480
\bibitem[Lee et al.(2003)]{Lee03}Lee, K. H., Lee, H. M., Fahlman, G. G., \& Lee, M. G.
    2003, \aj, 126, 815
\bibitem[Lee et al.(2007)]{Lee07}Lee, Y.-W., Gim, H. B., \& Casetti-Dinescu, D. I. 2007, \apj, 661, L49
\bibitem[Lee et al.(1999)]{Lee99}Lee, Y.-W., Joo, J.-M., Sohn, Y.-J., et al. 1999, \nat, 402, 55
\bibitem[Leon et al.(2000)]{Leo00}Leon, S., Meylan, G., \& Combes, F. 2000, \aap, 359, 907
\bibitem[Mackey \& Gilmore(2004)]{Mac04}Mackey, A. D., \& Gilmore. G. F. 2004, \mnras, 355, 504
\bibitem[Majewski et al.(2003)]{Maj03}Majewski, S. R., Skrutskie, M. F., Weinberg, M. D., \&
    Ostheimer, J. C. 2003, \apj, 599, 1082
\bibitem[Martin et al.(2004)]{Mart04}Martin, N. F., Ibata, R. A., Bellazzini, M., et al. 2004, \mnras, 348, 12
\bibitem[Martinez-Delgado et al.(2005)]{Mar05}Mart\'{i}nez-Delgado, D., Butler, D. J., Rix, H.-W., et al. 2005, \apj, 633, 205
\bibitem[Martinez-Delgado et al.(2004)]{Mar04}Mart\'{i}nez-Delgado, D., G\'{o}mez-Flechoso, M. \'{A}, 
	Aparicio, A., \& Carrera, R. 2004, \apj, 601, 242
\bibitem[McLaughlin \& van der Marel(2005)]{McLau05} McLaughlin, D.~E., \& van der Marel, R.~P.\ 2005, \apjs, 161, 304
\bibitem[McWilliam \& Rich(1994)]{McW94}McWilliam, A., \& Rich, R. M. 1994, \apjs, 91, 749
Vicari, A.\ 2006, \apj, 644, 940 
\bibitem[Montuori et al.(2007)]{Mon07} Montuori, M., Capuzzo-Dolcetta, R., Di Matteo, P., Lepinette, A., 
\& Miocchi, P.\ 2007, \apj, 659, 1212 
\bibitem[Moore et al.(2006)]{Moo06}Moore, B., Diemand, J., Madau, P., Zemp, M., \& Stadel, J.
    2006, \mnras, 368, 563
\bibitem[Moore et al.(1999)]{Moo99} Moore, B., Ghigna, S., Governato, F., et al. 1999, \apj, 524, L19
\bibitem[Nakasato \& Nomoto(2003)]{Nak03}Nakasato, N. \& Nomoto, K. 2003, \apj, 588, 842
\bibitem[Newberg et al.(2003)]{New03}Newberg, H. J., Yanny, B., Grebel, E. K., et al. 2003, \apj, 596, L191
\bibitem[Newberg et al.(2002)]{New02}Newberg, H. J., Yanny, B., Rockosi, C., et al. 2002, \apj, 569, 245
\bibitem[Newberg et al.(2009)]{New09}Newberg, H. J., Yanny, B., \& Willett, B. A. 2009, \apj, 700, L61
ed. H. Morrison \& A. Sarajedini (San Francisco: ASP), 14 
\bibitem[Noyola \& Gebhardt(2006)]{Noy06}Noyola, E., \& Gebhardt, K. 2006, \aj, 132, 447
\bibitem[Odenkirchen et al.(2003)]{Ode03}Odenkirchen, M., Grebel, E. K., Dehnen, W., et al. 2003, \aj, 126, 2385
\bibitem[Odenkirchen et al.(2009)]{Ode09}Odenkirchen, M., Grebel, E. K., Kayser, A., Rix, H.-W., \&
    Dehnen, W. 2009, \aj, 137, 3378
\bibitem[Odenkirchen et al.(2001)]{Ode01}Odenkirchen, M., Grebel, E. K., Rockosi, C. M., et al. 2001, \apj, 548, L165
\bibitem[Olszewski et al. (2009)]{Ols09}Olszewski, E. W., Saha, A., Knezek, P., et al. 2009, \aj, 138, 157
\bibitem[Possenti et al.(2003)]{Possenti2003} Possenti, A., D'Amico, N., Manchester, R.~N., et al.\ 2003, \apj, 599, 475
\bibitem[Rocha-Pinto et al.(2004)]{Roc04} Rocha-Pinto, H. J., Majewski, S. R., Skrutskie, M. F.,
     Crane, J. D., \& Patterson, R. J. 2004, \apj, 615, 732
\bibitem[Rockosi et al.(2002)]{Roc02}Rockosi, C. M., Odenkirchen, M., Grebel, E. K., et al. 2002, \aj, 124, 349
\bibitem[Schaerer \& Charbonnel(2011)]{Schaerer2011} Schaerer, D., \& Charbonnel, C.\ 2011, \mnras, 413, 2297
    2010, \apj, 725, 1175
\bibitem[Schlegel et al.(1998)]{Sch98}Schlegel, D. J., Finkbeiner, D. P., \& Davis, M.
    1998, \apj, 500, 525
\bibitem[Searle \& Zinn(1978)]{Searle1978} Searle, L., \& Zinn, R.\ 1978, \apj, 225, 357 
\bibitem[Shin et al.(2008)]{Shin2008} Shin, J., Kim, S.~S., 
\& Takahashi, K.\ 2008, \mnras, 386, L67
\bibitem[Sohn et al.(2003)]{Soh03}Sohn, Y.-J., Park, J.-H., Rey, S.-C., et al. 2003, \aj, 126, 803
\bibitem[Stetson(1987)]{Stetson1987}Stetson, P. B. 1987, \pasp, 99, 191
\bibitem[Stetson \& Harris(1988)]{Stetson1988}Stetson, P. B., \& Harris, W. E. 1988, \aj, 96, 909
\bibitem[Testa et al.(2000)]{Tes00}Testa, V., Zaggia, S. R., Andreon, S., et al. 2000, \aap, 356, 127
\bibitem[Trager et al.(1995)]{Tra95}Trager, S. C., King, Ivan R., \& Djorgovski, S. 1995, \aj, 109, 218
\bibitem[van den Bergh(1993)]{Van1993} van den Bergh, S.\ 1993, 
\apj, 411, 178
\bibitem[Vivas et al.(2001)]{Viv01}Vivas, A. K., Zinn, R., Andrews, P., et al. 2001, \apj, 554, L33
\bibitem[Vivas et al.(2008)]{Vivas2008} Vivas, A.~K., Jaff{\'e}, 
Y.~L., Zinn, R., et al.\ 2008, \aj, 136, 1645 
\bibitem[Vivas \& Zinn(2006)]{Vivas2006} Vivas, A.~K., \& Zinn, R.\ 2006, \aj, 132, 714
\bibitem[Wilson(1975)]{Wilson1975} Wilson, C.~P.\ 1975, \aj, 80, 175
\bibitem[Yanny et al.(2003)]{Yan03}Yanny, B., Newberg, H. J., Grebel, E. K., et al. 2003, \apj, 588, 824
\bibitem[Yanny et al.(2000)]{Yan00}Yanny, B., Newberg, H. J., Kent, S., et al. 2000, \apj, 540, 825
\bibitem[Zoccali et al.(2006)]{Zoc06}Zoccali, M., Lecureur, A., Barbuy, B., et al. 2006, \aap, 457, L1
\bibitem[Zoccali et al.(2003)]{Zoc03}Zoccali, M., Renzini, A., Ortolani, S., et al. 2003, \aap, 399, 931
\bibitem[Zucker et al.(2006)]{Zuc06}Zucker, D. B., Belokurov, V., Evans, N. W., et al. 2006, \apj, 650, L41


\end{thebibliography}
\end{document}